\documentstyle[prl,aps
]{revtex}
\begin{document}
\twocolumn[\hsize\textwidth\columnwidth\hsize\csname@twocolumnfalse%
\endcsname
\title{  
Quantal phases, disorder effects and superconductivity in spin-Peierls systems}
\author{Akihiro Tanaka and Xiao Hu}
\address{
Computational Materials Science Center, 
National Institute for Materials Science, Tsukuba 305-0047, Japan
}
\date{}
\maketitle
\begin{abstract}
In view of recent developments in the investigation on 
cuprate high-T${}_{\rm c}$ superconductors 
and the spin-Peierls compound CuGeO${}_{3}$, 
we study the effect of dilute impurity doping 
on the spin-Peierls state 
in quasi-one dimensional systems. 
We identify a common origin for 
the emergence of antiferromagnetic order upon 
the introduction of static vacancies, 
and superconductivity for mobile holes.

\end{abstract}
  ]  
\bigskip



The evolution of a disordered spin-gapped state 
(hereafter referred to as a spin liquid (SL)) into a superconductor 
as observed in the underdoped 
regime of the cuprate oxide compounds,  
continues to pose a major mystery. 
%
Here we address this problem from a new perspective; the comparison of 
the effect of static and mobile vacancies. 

Doping the CuO${}_2$ planes with static nonmagnetic impurities 
provides valuable information on the SLs.  
While monovalent Li can lead to additional complications\cite{Rice},   
Cu$\rightarrow$Zn substitution which introduces 
vacancies without excess holes serves as an ideal probe:   
enhancement of antiferromagnetic fluctuatuation accompanying the 
nucleation of local s=1/2 moments is observed \cite{Julien,Vojta},
apparently indicating that the state is a spin singlet with {\it
confined} spinon excitations.  Disorder-induced antiferromagnetism
(AF) is a property also shared among several quasi-1d spin-singlet
systems, 
most notably in the spin-Peierls compound
CuGeO${}_{3}$\cite{Regnault}, intensively studied  
in the past several years\cite{FTS}. 
Such analogy
has attracted much attention; the effect of disorder on the 
density of states of the staggered flux state 
has been discussed\cite{Pepin} in this light, 
as well as a possible explanation\cite{Fukuyama-Kohno-Keimer} 
of the extreme sensitivity  
of the 40 meV magnetic resonance peak to Zn impurities.   
However the precise  
relation, if any,  between 
such responses of SL against  ${\it static}$ impurities, 
and the superconducting instability observed in the cuprates 
(or spin-ladders)  
in the presence of 
${\it mobile}$ holes remains unclear. This is the 
issue we wish to take up below. 

In this article we have chosen to re-examine the simplest confining SL, 
the quasi-1d spin-Peierls (SP) system within the above scope. 
The primary aim is to 
invoke nonperturbative methods which 
takes advantage of the one-dimensionality, 
and to extract a physical picture which may well be generic to a wider family 
of confining SLs. Additional motivations come from experiments, however; 
while carrier doping 
so far has not been realized in CuGeO${}_3$,  
a possible proximity/coexistence of d-wave 
superconductivity with a SP-like (or bond-centered 
density wave) state in underdoped cuprates\cite{Fisher} 
has been inferred by very recent neutron scattering data of 
longitudinal optical phonon dispersions
in La${}_{1.85}$Sr${}_{0.15}$CuO${}_4$ and YBa${}_2$Cu${}_3$O${}_{6.95}$
\cite{Egami}.  

Let us summarize in physical terms what will follow. 
Theories on impurity effects in spin chains 
have often focused on either the formation of local spin moments   
(using AB), or on quantum interferences among 
a preassigned array of antiferromagnetically aligned spins in  
the presence of a few vacant sites (employing  
semiclassical methods).  
The former cannot account for directional fluctuations of the spins, 
while the latter may overemphasize the role of local spin moments 
by assuming them even in a spin singlet state. 
Our remedy is to devise a version of nonabelian bosonization (NAB) which 
simultaneously resolves both aspects.  
We derive an effective 
action reproducing the 
known AB theory for SP systems\cite{Nakano-Inagaki-Fukuyama}, 
{\it with} two additional terms,  
each related to the directional fluctuations of the AF order parameter  
and the Berry phases. The contribution of these new terms are appreciable 
only near an impurity-induced spin moment. This local weight transfer 
together with the Berry phase effect (which keeps the spin moments 
in registry with the underlying AF pattern) 
is what gives rise to AF order in the {\it static}  
impurity case. 
%
An important aspect here is the length (or energy) scales involved. 
Adiabatic adjustment of the spins to the 
charge deficiency requires the healing length 
of the charge to be sufficiently shorter than that for the spins.    
Meanwhile when {\it mobile} holes are doped, 
the same basic conditions are seen to 
enhance superconducting instability, enabling us to make contact with 
a pairing picture 
proposed  
in several earlier works. In this case, 
the local AF enviroment provided by the spectral weight transfer
enhances intrasublattice hopping as well as mediates intersublattice attraction. 
The adiabadicity condition 
ensures slow fluctuation of the AF enviroment, 
necesarry for coherent 
motion of the holes. 

To model a SP system, we incorporate 
the one-band Peierls-Hubbard (PH) model at 
half-filling, 
\begin{equation}
{\rm H}=\sum_{i\sigma}(t-(-1)^{i}\delta t)
(c^{\dagger}_{i\sigma}c_{i\sigma}
+{\rm h.c.})+U\sum_{i}n_{i\uparrow} n_{i\downarrow}, 
\label{PH model1}
\end{equation}
with $U>0$. 
Interchain coupling (which 
fixes the preference of the dimer pattern and 
hence confines disorder-induced spinons) will be assumed. 
We start by a semiclassical description of the bulk state; 
treating the $U$-term as a 
commensurate spin density wave (cSDW) saddle point solution 
${\vec \varphi}=<c^{\dagger}
\frac{\vec \sigma}{2}c>$, with  
${\vec \varphi}_{i}=(-1)^{i}
m{\vec n}_{i}$ ($\vert {\vec n}_{i}\vert=1$, 
$m\sim\frac{t}{U}e^{-\frac{6\pi t}{U}}$\cite{Affleck Les Houches}), 
we get a  4-component dirac-fermion type Hamiltonian density\cite{AM} 
\begin{equation} 
{\cal H}_{\rm F}=
\left[R^{\dagger}, L^ {\dagger} 
\right]
\left[
\begin{array}{c}
 -iv_F 
\partial_{x}, 
-\Delta_0 
Q
{\rm e}^{-{\rm i}
Q\frac{\phi_0}{2}} \\ 
-\Delta_0 
Q{\rm e}^{{\rm i}
Q\frac{\phi_0}{2}},
 iv_F 
\partial_x
\end{array}
\right]
\left[
\begin{array}{c}
 R \\ 
 L
\end{array}
\right],
\label{continuum Hamiltonian}
\end{equation}
where the right ($R$) and left ($L$) movers each carry a spin index, 
$Q\equiv{\vec n}\cdot{\vec \sigma}$, $\Delta_{0}
\equiv\sqrt{(\frac{4Um}{3})^{2}+(2\delta t)^{2}}$, and ${\rm
tan}(\frac{\phi_0}{2})\equiv\frac{3\delta t}{2Um}$.  
%
To see the physics embodied in 
the mass term 
(off-diagonal elements) 
assume temporarily that ${\vec n}\equiv {\hat z}$.  
The effective spin-dependent potential energy  
$V_{\sigma}(x)\equiv
e^{i2k_F x}V_{\sigma}(2k_F)+
e^{-i2k_F x}V_{\sigma}(-2k_F)$ 
experienced by each spin component $\sigma=\pm 1$ can be read off 
using 
${\cal H}_{\rm off-diag}=
\sum_{\sigma}R_{\sigma}^{\dagger}L_{\sigma}V_{\sigma}(2k_{F})+
L_{\sigma}^{\dagger}R_{\sigma}V_{\sigma}(-2k_{F})$:  
\begin{equation}
V_{\sigma}(x)=\left\{
 \begin{array}{rl}
  2\Delta_0 \sin(2k_F x
+(\frac{\pi}{2}+\frac{\phi_0}{2}))& (\sigma=
+1).\\
 2\Delta_0 
\sin(2k_F x-(\frac{\pi}{2}+\frac{\phi_0}{2}))
& (\sigma=
-1).
\end{array}\right.
\label{intuitive1}
\end{equation}
The minima of the potential are located at $x_{min}={\rm X}a$ 
($a$: lattice constant), where 
${\rm X}=
(2n-1)-\frac{\phi_0}{2\pi},n\in {\bf Z}$ for $\sigma=
+1$ and  ${\rm X}=
2n+\frac{\phi_0}{2\pi},n\in {\bf Z}$ for $\sigma=
-1$, which 
invites the following interpretation. 
When $\phi_0=0$, odd sites (even sites) are 
occupied by down-spins (up-spins)
(the cSDW theory). 
Turning on electron-lattice coupling ($\phi_0 \ne 0$) 
shifts the position of 
these down-spins (up-spins) to the 
right (the left) resulting in a 
regular array of strong (odd-even) and weak 
(even-odd) bonds.  
The antiparallel spin pairs on the strong bonds   
form 
bond-centered 
density waves. 
Returning to the general case, 
this picture remains locally valid 
with the replacement 
${\hat z}\rightarrow{\vec n}$  
provided ${\vec n}$ fluctuates on a scale longer than 
$a$. 

A low energy theory is obtained by 
treating the corresponding Lagrangian   
${\cal L}_{\rm F}={\bar \Psi}
\left[
{\bf 1}\otimes
{\not\! \partial}
-\Delta_{0}Q e^{iQ\frac{\phi_0}{2}\gamma^5}
\right]
\Psi$ 
(where 
${\bar \Psi}\equiv\Psi^{\dagger}\gamma_{0}$, 
$\gamma^{5}\equiv i\gamma_0 \gamma_1$, and  
$e^{iQ\frac{\phi_0}{2}\gamma^5 }\equiv
\cos\frac{\phi_0}{2}{\bf 1}\otimes{\bf 1}+
i\sin\frac{\phi_0}{2}Q\otimes\gamma^5$) in a 
derivative expansion of ${\vec n}$. 
Perturbative terms enter only 
beyond quadratic order\cite{Tsvelik book}, and the resulting action - coming from the 
anomaly of the SU(2) current  
${\bf j}_{\mu}^{5}\equiv{\bar \Psi}\gamma_{\mu}\gamma^{5}\frac{\vec \sigma}{2}\Psi$ -  
is the O(3) nonlinear sigma 
(NL$\sigma$) model with vacuum angle $\theta=\pi-\phi_0 -\sin\phi_0$\cite{AM,Abanov Wiegmann}, which signals a 
spin-gap for $\phi_0 \ne 0$. 

Having discussed the bulk system at 1/2-filling, 
we now introduce a dilute density of 
nonmagnetic impurities 
(vacancies). 
To this end, keeping $\phi_0$ constant 
(justified in the presence of interchain coupling), 
we bosonize. 
In doing so we must go beyond usual AB with a 
fixed spin quantization axis
in order to retain the spin directional degree of freedom ${\vec n}$. 
We describe in some detail how this can be done. 
The AB scheme derives from 
the fermion-boson correspondence 
\begin{eqnarray}
R_{\sigma}&\propto&
e^{\frac{i}{2}[\theta_{+}+\theta_{-}+\sigma
(\phi_{+}+\phi_{-})]}\nonumber\\
L_{\sigma}&\propto&
e^{\frac{i}{2}[-\theta_{+}+\theta_{-}+\sigma
(-\phi_{+}+\phi_{-})]},
\label{fermion to boson}
\end{eqnarray}
(in conventions of 
ref.\cite{bosonization}), and translates 
fermionic operators into the language of  
a set of conjugate charge ($\theta_{\pm}$) and  
spin ($\phi_{\pm}$) phase fields.  
We attempt a modification by the simple 
replacement $\sigma\rightarrow Q={\vec n}\cdot{\vec \sigma}$. 
Constructing fermi-bilinears according to this rule, 
one readily sees that it corresponds to parametrising the 
k=1 Wess-Zumino-Witten (WZW) field $g\in$SU(2)  
appearing in the NAB rules\cite{Tsvelik book} as 
$g=e^{-i\phi_{+}Q}$. (The dual fields $\phi_{-}$ and $\theta_{-}$ 
are gauge degrees of freedom which do not arise unless 
considering chiral currents or Cooper channels). 
Note that we differ from the 
usual way\cite{Affleck Les Houches} 
of relating abelian phase fields to the WZW model. 
To give firmer grounds to this identification observe that 
eq.(\ref{fermion to boson}) and its nonabelian generalization  
may be viewed as a family of chiral transformations 
acting on a bosonic vacuum free of charge or spin solitons.   
The spin part of the free fermion theory therefore  
has an induced SU(2) connection and is consistently evaluated as 
\begin{eqnarray}
Z_{spin}&=&
\int{\cal D}{\vec n}{\cal D}{\phi_{+}}
\int{\cal D}{\bar \Psi}{\cal D}\Psi 
e^{-\int d\tau dx {\bar \Psi}[{\bf 1}\otimes
{\not\partial}+U_5{\not\partial} U_5]\Psi}
\nonumber\\
&=&
\int{\cal D}{\vec n}{\cal D}{\phi_{+}}
e^{-S_{\rm wzw}[g]}\vert_{g=e^{-i\phi_{+}Q}}=Z_{\rm wzw},
\end{eqnarray} 
where $U_{5}\equiv e^{-\frac{i}{2}\phi_{+}Q\gamma^{5}}$. 
The bosonization dictionary 
remains unaltered for the charge sector, while 
the spin part 
receives corrections related to the fluctuations of  
${\vec n}$. For instance the $k=2k_F$ component of 
the spin operator reads  
${\vec S}_{k=2k_{F}}\propto\sin(2k_F x+\theta_{+})
\sin\phi_{+}{\vec n}$, whereas the uniform component is
\begin{equation}
{\vec S}_{k=0}={\bf J}_{\rm R}+{\bf J}_{\rm L}=\frac{1}{2\pi}\partial_{x}\phi_{+}
{\vec n}+{\vec S}_{\rm additional}
\label{k=0 component}
\end{equation}
in which  
${\vec S}_{\rm additional}=
\frac{1}{2\pi}\cos\phi_{+}\sin\phi_{+}\partial_{x}{\vec n}
-\frac{1}{2\pi}\sin^{2}\phi_{+}{\vec n}\times
\partial_{x}{\vec n}$. 
Canonical quantization of ${\vec n}$ and $\phi_{+}$ yields the correct (Kac-Moody) algebra 
for the currents ${\bf J}_{\rm R}$ and ${\bf J}_{\rm L}$.  
The bosonized Lagrangian for ${\cal L}_{F}$ is 
${\cal L}={\cal L}(\theta_{+})
+{\cal L}_{\rm wzw}(g)\vert_{g=\exp(-i\phi_{+}Q)}+
{\cal L}_{mass}(\theta_{+},\phi_{+},{\vec n})$,
where the free fermion parts for the charge and spin are 
respectively 
${\cal L}_{\theta_{+}}
=\frac{1}{4\pi}(\partial_{\mu}\theta_{+})^{2}$ and 
\begin{eqnarray} 
{\cal L}_{\rm wzw}(g)\vert_{g=\exp(-i\phi_{+}Q)}
&=&
\frac{1}{4\pi}(\partial_{\mu}\phi_{+})^{2}
+\frac{1}{8\pi}\sin^2{\phi_{+}}
(\partial_{\mu}{\vec n})^{2}
\nonumber\\
&&
+i(2\phi_{+}-\sin (2\phi_{+}))q_{\tau x},
\label{wzw-part}
\end{eqnarray}
with 
$q_{\tau x}=\frac{1}{4\pi}
{\vec n}\cdot{\partial}_{\tau}{\vec n}\times
\partial_{x}{\vec n}$, the topological charge density 
of instantons. The last two terms in eq.(\ref{wzw-part}) are 
new; they are terms that disappear on taking ${\vec n}$=const., 
reproducing the 
AB expression for a free Tomonaga-Luttinger liquid. 
On the other hand, 
a bulk spin gap  
will fix $\phi_{+}$, and (for $\phi_{+}\ne 0$) 
these terms will 
yield an O(3) NL$\sigma$ model with a $\theta$-term. 
The mean value of $\phi_{+}$ 
is determined from the interaction ${\cal L}_{mass}$, 
but the latter needs be handled with some care. 
Straightforward bosonization 
gives 
\begin{equation}
{\cal L}_{mass}=\frac{2}{\pi\alpha}\sin\theta_{+}
\cos(\phi_{+}-\frac{\pi}{2}+\frac{\phi_0}{2})
\label{uncorrected mass term}
\end{equation}
where $\alpha$ is a 
short distance 
cutoff. Being a relevant term, this locks 
$\phi_{+}$ in the bulk problem to the value 
$\phi_{+}=\frac{\pi}{2}-\frac{\phi_0}{2}$. (We assume that  
$\sin\theta_{+}$ attains a mean value, i.e. 
the Umklapp term 
present at half-filling opens up the charge gap.) 
Plugging this into 
${\cal L}_{\rm wzw}$, we recover the NL$\sigma$ model with 
the $\theta$-angle previously mentioned. 
But to see the full correspondence with the AB result, 
we should go beyond this 
semiclassical approximation; noting that the portion of the mass term 
$\Delta_{0}\cos\frac{\phi}{2}(R_{\alpha}^{\dagger}Q_{\alpha\beta}L_{\beta}+
h.c.)$ had originated (prior to the decoupling) 
from a backscattering process, 
we should correct eq.(\ref{uncorrected mass term}) into the form 
\begin{equation}
{\cal L}_{mass}=\frac{2\Delta_{0}}{\pi\alpha}\sin\theta_{+}
\sin(\frac{\phi_0}{2})\cos\phi_{+}+D\cos 2\phi_{+},  
\label{corrected mass term}
\end{equation}
where we now have complete agreement with the well-known  
AB result for the spin-Peierls system
\cite{Nakano-Inagaki-Fukuyama}, supplemented with the 
second and third terms of eq.(\ref{wzw-part}).
The $D$-term is marginally irrelevant, and the effective 
value of $\theta$ 
is now governed by the first term of 
eq.(\ref{corrected mass term}). For 
$\phi_0\ne 0 
$ this gives  
$\phi_{+}=0$ and hence $\theta=0$, in which 
case the magnitude of the staggered spin $m{\vec n}$
is quenched, 
making the second term in eq.(\ref{wzw-part}) 
ineffective. 
(The formula for ${\vec S}_{k=2k_{F}}$ infers that 
$\theta=0$ and $\pi$ each corresponds to a spin-singlet and 
a N\'eel state.) 
The case $\phi_0=0$ (no dimerization) is special;  
only the $D$-term is present and the effective 
$\theta$-angle is undetermined, indicating a dynamically  
induced axial U(1) symmetry. 
This suggests a physical picture of fluctuating dimers reminiscent 
of a long-ranged resonating-valence-bond state 
\cite{Nakano-Inagaki-Fukuyama}.  
The above arguments expose an 
intimite and rather unexpected relation between the $\theta$-angle and the 
spin phase fields of AB, which only becomes evident in the present 
^^ ^^ rotating frame''. 
In the remaining part 
we seek its consequences,  
concentrating on the case 
$\phi_0\ne 0$. 

We are now ready to discuss vacancies. 
From the formula for the charge density  
$\rho=\frac{1}{\pi}\partial_{x}\theta_{+}$, 
this is represented by a 
$\pi$-kink of $\theta_{+}$. From either 
eq.(\ref{uncorrected mass term}) or 
eq.(\ref{corrected mass term}), this is seen to 
invert the sign of the 
potential energy, which must be compensated by a 
$\pi$-kink of $\phi_{+}$. The latter corresponds, according to 
eq.(\ref{k=0 component}) to the liberation of a spin 1/2 degree of 
freedom in the background of the singlet state.
({\it Note that this argument does not 
apply in the absence of the spin gap, i.e. for $\phi_0=0$.}) 
It is possible to show, 
that along the lines of ref\cite{Saito-thesis} the spectral 
weight Im$\chi(k,\omega)$ for fixed $\phi_{+}$ can be estimated as 
$\sim\frac{\cos^{2}\phi_{+}}{\sqrt{(k-\pi)^2 +m^2}}$ for the 
gapped part, while another contribution 
$\sim\vert\sin\phi_{+}/(k-\pi) \vert$ represents a spin wave-like part. 
Hence the $\pi$-kink of $\phi_{+}$ should indeed cause a transfer of 
spectral weight into subgap states, which is a characteristic feature 
of the present system. 
Physically, a vacancy at site $x=X_{i}$ should 
release a spin of ${\vec S}\sim\frac{1}{2}
(-1)^{X_{i}/a}{\vec n}(X_{i})$, which 
fixes the sign of the kink of the $\phi_{+}$ field to be 
$\delta\phi_{+}=\pi\sum_{X_{i}}(-1)^{X_{i}/a} \Theta(x-X_{i})
=(-1)^{X_{i}/a}\theta_{+}$. ($\Theta$ is the step function.) 
This should be true under the adiabadicity condition 
$\xi_c \ll\xi_s$ where $\xi_c$ ($\xi_s$) is the 
charge (spin) correlation length. 
The continuum limit however does not 
distinguish which of the two sublattices a given point $x=X_{i}$ 
belongs to. Such lattice effects can be 
particularly important when dealing with Berry phases\cite{Sachdev-Park}.  
To continue working in the continuum, 
we are thus lead to introduce {\it two} 
charge phase fields $\theta_{+}^{A}$ and 
$\theta_{+}^{B}$, one for each sublattice. 
This leads to a 
simple expression for the deviation 
$\delta\phi_{+}$ of $\phi_{+}$ from the bulk value $\bar{\phi}_{+}$, 
\begin{equation}
\delta\phi_{+}=\theta_{+}^{A}-\theta_{+}^{B},
\label{pri}
\end{equation}
which is the principal equation of this article.  
Now let us see how this affects the topological term, 
${\cal L}_{top}=i[2\phi_{+}-\sin(2\phi_{+})]q_{\tau x}$. 
Again using $\xi_c \ll\xi_s$, the effect of $\delta\phi_{+}$ on the 
term $-i\sin(2\phi_{+})q_{\tau x}$ cancels out on average, and  
\begin{equation}
{\cal L}_{top}
=i[2\bar{\phi}_{+}-\sin 2\bar{\phi}_{+}]q_{\tau x}
+2i[\theta_{+}^{A}-\theta_{+}^{B}]q_{\tau x},
\label{top-term1}
\end{equation}
provided the average seperation 
of vacancies $l>\xi_s$. 
Next, 
we note\cite{Haldane PRL} that formally 
$q_{\tau x}=\frac{1}{4\pi}\partial_{x}A_{0}$, where 
$A_{0}(\tau,x)\equiv \partial_{\tau}{\vec n}\cdot
{\vec a}({\vec n}(\tau,x))$, 
and the monopole vector potential ${\vec a}$ satisfies 
$\nabla_{\vec n}\times{\vec a}={\vec n}$. Integrating by parts, the 
second term in eq.(\ref{top-term1}) becomes 
${\cal L}_{top}'=-\frac{i}{2\pi}\partial_{x}
(\theta_{+}^{A}-\theta_{+}^{B}){\vec a}\cdot\partial_{\tau}
{\vec n}$. For static vacancies, this yields the action 
$S_{top}'=\frac{i}{2}\sum_{X_{i}}(-1)^{X_{i}/a}
\omega[{\vec n}(\tau, X_{i})]$, where 
$\omega[{\vec n}(\tau)]=\int d\tau A_{0}$ is the solid angle 
subtended by ${\vec n}(\tau)$ in the course of its evolution. 
These terms are  
the Berry phases of spin 1/2 objects induced by vacancies. 
Together with the bulk contributions consisting of the NL$\sigma$ model 
and the first term of eq.(\ref{top-term1}) (we use for the potential 
eq.(\ref{uncorrected mass term})), this is essentially the 
action derived in ref.\cite{Nagaosa ladder} for the doping of a 
spin ladder. Following similar arguments, we arrive at the 
final action for the induced spins $\{{\vec n}(X_{i})\}$, 
\begin{eqnarray}
S_{eff}[\{{\bf n}_{j}\}]
&=&\sum_{j}\frac{i}{2}(-1)^{X_{j}/a}\omega[{\bf n}_{j}(\tau)]
\nonumber\\
&&-\int d\tau J_{eff}e^{-\vert X_{j}-X_{j+1}\vert/\xi}
{\bf n}_{j}(\tau)\cdot{\bf n}_{j+1}(\tau),
\end{eqnarray}
where ${\bf n}_{j}\equiv {\vec n}({X_j})$ and 
$J_{eff}=1/\xi\cdot \sin^{2}\frac{\phi_0}{2}$. 
Absorbing the signs into the spins 
${\bf N}_{j}\equiv(-1)^{X_{j}/a}{\bf n}_{j}$, this becomes a random exchange 
Heisenberg model, with diverging spin correlation and 
staggered susceptibility at T=0\cite{Nagaosa ladder}. 
We expect that the essential physics of the 
disorder-induced AF 
observed in CuGeO$_{3}$ is captured within this model. 

Turning to the case of mobile vacancies (holes), 
the part of the action involving 
$\theta_{+}^{A}$ and $\theta_{+}^{B}$ reads 
\begin{eqnarray}
{\cal L}
(\theta_{+}^{A},\theta_{+}^{B})
&=&\frac{1}{8\pi}\left[\partial_{\mu}
(\theta_{+}^{A}+\theta_{+}^{B})\right]^{2}
+\frac{1}{8\pi}\left[\partial_{\mu}(\theta_{+}^{A}-\theta_{+}^{B})
\right]^{2}
\nonumber\\
&&+2i[\theta_{+}^{A}-\theta_{+}^{B}]q_{\tau x}.
\end{eqnarray}
This coincides with the action proposed by 
Shankar\cite{Shankar} on semiphenomenological 
grounds for hole motions in an 
antiferromagnetic background. We have arrived at this form 
from an electron system containing 
both spin and charge sectors.  
Hereon we may basically adapt the arguments of ref.\cite{Shankar}. 
Refermionizing ${\cal L}(\theta_{+}^{A},\theta_{+}^{B})$ we  
see that it is equivalent to two massless fermions 
\begin{equation}
{\cal L}_{hole}=
{\bar \psi}_{A}({\not\! \partial}+i{\not\! A})\psi_{A}  
+{\bar \psi}_{B}({\not\! \partial}-i{\not\! A})\psi_{B}.
\end{equation}
coupled to the gauge fields 
$A_{\mu}={\vec n}\cdot\partial_{\mu}{\vec a}$ 
each describing intrasublattice 
(next nearest neighbor) hopping of the holes\cite{Shankar,Auerbach book}. 
Because the fermions $\psi_{A}$ and $\psi_{B}$ have opposite gauge charges, 
their is an attractive interaction. 
The spin singlet superconducting susceptibility is the 
correlation fuction of (in terms of the original electrons)
$\psi_{R\sigma}^{A}(x){\psi^{B}}_{ L-\sigma}(x)\sim
e^{\frac{i}{2}(\theta_{+}^A - \theta_{+}^B )}
e^{\frac{i}{2}(\theta_{-}^A +\theta_{-}^B)}e^{i\phi_{+}}$.  
The nontrivial combination here is $\theta_{+}^{A}-\theta_{+}^{B}$, 
but Gauss law constraints can be incorporated \cite{Shankar} to show it is 
massive.  
Then the susceptibility obeys a power law with  
exponent -1, and should become an Emery-Luther superconductor when 
including interactions among neighboring sites, 
which seems to be consistent with available 
numerical results on dimerized t-J models\cite{Imada}. 
This would suggest the possibility of the 
coexistence of the spin-gapped state (i.e. spin-Peierls state) with 
superconductivity. Superconductivity in quasi-1d spin-gapped systems 
may become relevant in view of the 
recent advances in hole-injection via field-effect transistors\cite{Schon}. 
Finally we note certain differences from ref\cite{Shankar}; 
first the picture relies on a confining SL, therefore breaking 
down at the Heisenberg point $\phi_0 =0$, i.e.  
a spin gap is required. A second feature is the difference in the 
vacua structure; 
we had effectively $\theta=0$ irrespective of the value of 
$\phi_0\ne 0$, and hence zero weight 
for the NL$\sigma$ model part, so the collapse of a $\theta$-vacua 
structure with hole doping\cite{Shankar} is not seen, which marks a 
departure from semiclassical methods. 

The crux of our argument is that the enhancement of superconducting  
susceptibility, 
related to the pairing scenarios  
based on t'-J type interactions\cite{Auerbach book}, 
has emerged here from the same origin as the disorder-induced AF, 
which is the coupling of the charge density fluctuation to the 
spin gauge fluctuation, represented compactly in 
eq.(\ref{pri}).
This suggests  
 that the validity of such pictures 
in a particular spin-gapped system 
can be inferred from testing its response to static nonmagnetic impurities. 
In this respect 
we mention recent hole-doping experiments in the Haldane-gap material  
Y${}_2$BaNiO${}_5$\cite{Ito}, a system where 
static impurities do not induce AF.  
The authors find no enhancement of conductivity and ascribe it to the 
comparable magnitude of the charge and spin gaps. 

Since our framework can be readily applied to charge stripes 
modeled as arrays of 1d electron gases
 coupled to spin-gapped chains\cite{stripe1}, 
similar treatments should provide useful insight. 
Such work is now in progress.

We wish to thank M. Saito and 
H. Fukuyama for sharing with us  
their insights on disordered spin-Peierls systems. 
We acknowledge 
M. Hase, K. Uchinokura, N. Taniguchi and T. Hikihara
for many discussions, and I. Affleck for several helpful comments 
on the occasion of the conference RPMBT11.

\end{document}